
%
\input harvmac
\noblackbox
%
%

\def\apm{\alpha^{\prime}}

%
%
\lref\guven{R. G\"{u}ven, ``Black $p$-brane solutions of $D=11$
supergravity theory,'' Phys. Lett. {\bf B276} (1992) 49.}
\lref\sdualref{A. Font, L. Ibanez, D. Lust, and F. Quevedo, Phys. Lett. {\bf
249B}
(1990) 35; S. J. Rey, Phys. Rev. {\bf D43} (1991) 526;
 (1992) 374; J. Harvey, J. Gauntlett and J. Liu, Nucl. Phys. {\bf 409B } (1993)
 363; R. Khuri, Phys. Lett. {\bf 259B} (1991) 261;  J.H. Schwarz and A. Sen,
Nucl. Phys. {\bf 404B }
(1993) 109;
J. Gauntlett and J. Harvey, hep-th/9407111; A. Sen, Int. J. Mod. Phys. {\bf A9}
(1994)
3707, hep-th/9402002; J. H. Schwarz, hep-th/9411178, C. Vafa and E. Witten,
hep-th/9408074; Nucl. Phys. {\bf B431} (1994) 3, L. Girardello, A. Giveon, M.
Porrati
and A. Zaffaroni, Phys.Lett. {\bf B334} (1994) 331, hep-th/9406128;
J. Harvey, G. Moore and A. Strominger, hep-th/9501022;  M. Bershadsky, A.
Johansen, V. Sadov and C. Vafa, hep-th/9501096.}
\lref\wit{E. Witten, ``String Theory Dynamics in Various Dimensions,''
hep-th/9503124.}
\lref\rey{S. J. Rey, Phys. Rev. {\bf D43} (1991) 526.}
\lref\drev{For a review see M. J. Duff,  R. R. Khuri, and J. X. Lu, ``String
Solitons,'' hep-th/9412184.}
\lref\senone{A. Sen, ``Macroscopic Charged Heterotic String'', hep-th/9206016,
Nucl. Phys. {\bf B388} (1992) 457.}
\lref\sentwo{A. Sen, ``String-String Duality Conjecture in Six Dimensions and
Charged Solitonic Strings'', hep-th/9504027.}
\lref\nati{N. Seiberg, ``Observations on the Moduli Space of Superconformal
Field Theories, ''Nucl. Phys. {\bf B303}, (1986), 288;  P. Aspinwall and D.
Morrison,  ``String
Theory on K3 Surfaces,'' preprint DUK-TH-94-68, IASSNS-HEP-94/23,
hep-th/9404151.}
\lref\crdy{J. Cardy and E. Rabinovici, Nucl. Phys. {\bf B205} (1982) 1;
J. Cardy, Nucl. Phys. {\bf B205} (1982) 17.}
\lref\filq{A. Font, L. Ibanez, D. Lust, and F. Quevedo, Phys. Lett. {\bf 249B}
(1990) 35.}
\lref\rey{S. J. Rey, Phys. Rev. {\bf D43} (1991) 526.}
\lref\sen{A. Sen, Int.J.Mod.Phys. {\bf A9}  (1994) 3707,  hep-th/9402002.}
\lref\sentform{A. Sen,
Phys. Lett. {\bf B329} (1994) 217, hep-th/9402032.}
\lref\wol{E. Witten and D. Olive, Phys. Lett. {\bf 78B} (1978) 97.}
\lref\osb{H. Osborne,
``Topological Charges for N=4 Supersymmetric
Gauge Theories and Monopoles of Spin 1, ''
Phys. Lett.
{\bf 83B}(1979)321.}
\lref\mol{C. Montonen and D. Olive, Phys. Lett. {\bf 72B} (1977) 117.}
\lref\swit{N. Seiberg and E. Witten, ``Electromagnetic Duality,
Monopole Condensation and Confinement in N=2 Supersymmetric Yang-Mills Theory,
''
hep-th/9407087.}
\lref\vafa{C. Vafa, as referenced in  \wit.}
\lref\ferr{A. Ceresole, R. D'Auria, S. Ferrara and A. Van Proeyen,
``Duality Transformations in Supersymmetric Yang-Mills Theory Coupled
to Supergravity, '' hep-th/9502072.}
\lref\schsen{J. Schwarz and A. Sen, Phys. Lett. {\bf B312} (1993) 105,
hepth/9305185; P. Binetruy, Phys. Lett. {\bf B315} (1993) 80, hep-th/9305069. }
\lref\jpas{J. Polchinski and A. Strominger,  ``Effective String Theory, ''
Phys. Rev. Lett. {\bf 67} (1991) 1681.}
\lref\wz{Y. S. Wu and A. Zee,  `` A Closed String (or Ring) Soliton
Configuration with Nonzero Hopf Number, '' Nucl. Phys. {\bf B324}
(1989) 623. }
\lref\ddp{J. A. Dixon, M. J. Duff and J. C Plefka, Phys. Rev. Lett.
{\bf 69} (1992) 3009.}
\lref\izt{J. M. Izquierdo and P. K. Townsend,  ``Axionic Defect Anomalies and
their
Cancellation, '' Nucl. Phys. {\bf B414} (1994) 93,  hepth/9307050.}
\lref\jb{J. Blum and J. A. Harvey, ``Anomaly Inflow for Gauge Defects, ''
Nucl. Phys. {\bf B416} (1994) 119, hepth/9310035.}
\lref\DAHA{ A. Dabholkar and J. A. Harvey,  ``Nonrenormalization of
the Superstring Tension, '' Phys. Rev. Lett.
{\bf 63} (1989) 478.}
\lref\TEN{ I. C. G. Campbell and P. West, Nucl. Phys. {\bf B243},
(1984) 112; F. Giani and M. Pernici, Phys. Rev. {\bf D30}, (1984)
325;
M. Huq and M. A. Namazie, Class. Quant. Grav. {\bf 2} (1985) 293.}
\lref\GSW{M. B. Green, J. H. Schwarz, and E. Witten, {\it Superstring
Theory} , Vol. 2. ,
Cambridge University Press (1987).}
\lref\gau{J. Gauntlett and J. A. Harvey,  ``S-duality and the Spectrum of
Magnetic Monopoles
in Heterotic String Theory, '' hep-th/9407111.}
\lref\DGHR{ A. Dabholkar, G. Gibbons, J. A. Harvey and F. R. Ruiz,
``Superstrings and Solitons, '' Nucl. Phys. {\bf B340} (1990) 33.}
\lref\duffive{ M. J Duff,  ``Supermembranes, the First Fifteen Weeks, ''
Class. Quant. Grav. {\bf 5} (1988) 189.}
\lref\hetsol{ A. Strominger,  ``Heterotic Solitons, ''
Nucl. Phys. {\bf B343} (1990) 167; E: Nucl. Phys. {\bf B353} (1991) 565.}
\lref\HASTR{ J. A. Harvey and A. Strominger, ``Octonionic Superstring
Solitons, '' Phys. Rev. Lett. {\bf 66} (1991) 549.}
\lref\ADHM{M. F. Atiyah, V. G. Drinfeld, N. J. Hitchin and Y. I. Manin,
``Construction of Instantons,'' Phys. Lett. {\bf 65A} (1978) 185.}
\lref\hoof{See e.g. R. Rajaraman, {\it Solitons and Instantons}, North-Holland,
1982.}
\lref\sswt{ A. Strominger,  ``Superstrings with Torsion '',
Nucl. Phys. {\bf B274} (1986) 253.}
\lref\towne{P.K.  Townsend,  ``The Eleven-Dimensional Supermembrane
Revisited,''   R/95/2,
hep-th/9501068. }
\lref\townrev{P.K.  Townsend,  ``Three Lectures on Supermembranes'' in
{\it Superstrings '88} eds. M. Green,
M. Grisaru, R. Iengo, E. Sezgin and A. Strominger, World Scientific, Singapore,
1989.}
\lref\town{C.M. Hull and P.K.  Townsend, ``Unity of Superstring Dualites,''
QMW-94-30, R/94/33,
hep-th/9410167. }
\lref\polc{J.  Dai, R. G. Leigh, and J. Polchinski,  ``New Connections Between
String
Theories,'' Mod. Phys. Lett. {\bf A4} (1989) 2073. }
\lref\dhs{M. Dine, P. Huet, and N. Seiberg, ``Large and Small Radius in String
Theory,''
Nucl. Phys. {\bf B322} (1989) 2073. }
\lref\RW{ R. Rohm and E. Witten, ``The Anti-Symmetric Tensor Field in
Superstring Theory,'' Ann. Phys. {\bf 170}, 454 (1986). }
\lref\STR{ A. Strominger, ``Superstrings with Torsion, '' Nuclear
Physics {\bf B274}, 253 (1986). }
\lref\duffone{M. J. Duff and J. X. Lu,
``Elementary Fivebrane Solutions of D=10
Supergravity,'' Nucl. Phys. {\bf B354} (1991) 141.}
\lref\dufftwo{M. J Duff and J. X. Lu,  ``Remarks on String/Fivebrane
Duality,'' Nucl. Phys. {\bf B354} (1991) 129.}
\lref\duffthree{M. J. Duff and J. X. Lu, ``Strings from Fivebranes,''
Phys. Rev. Lett. {\bf 66} (1991) 1402.}
\lref\MOOL{ C. Montonen and D. Olive, ``Magnetic
Monopoles as Gauge Particles? '' Phys. Lett. {\bf 72B} (1977) 117. }
\lref\hupo {J. Hughes and J. Polchinski, Nucl. Phys. B{\bf278}, 147 (1986).}
\lref\TOW{ P. K. Towsend ``Supersymmetric Extended Solitons,''
 Phys. Lett. {\bf 202B} (1988) 53. }
\lref\bhole{G. Horowitz and A. Strominger, ``Black Strings and
$p$-branes,'' Nucl. Phys. {\bf B360} (1991) 197.}
\lref\wbran{C. G. Callan, J. A. Harvey and A. Strominger,
``Worldbrane Actions for String Solitons,'' Nucl. Phys. {\bf B367} (1991) 60.}
\lref\tei{R. Nepomechie, Phys. Rev. {\bf D31} (1985) 1921;
C. Teitelboim, Phys. Lett. {\bf B176} (1986) 69.}
\lref\wsheet{C.G. Callan, J. A. Harvey and A. Strominger,
``Worldsheet Approach to Heterotic Instantons and  Solitons,'' Nucl. Phys. {\bf
359}
(1991) 611. }
\lref\no{H. Nielsen and P. Olesen, Nucl. Phys. {\bf B61} (1973) 45. }
\lref\duffcq{M. J. Duff and J. X. Lu,  ``Remarks on String/Fivebrane
Duality,'' Nucl. Phys. {\bf B354} (1991) 129; ``String/Fivebrane Duality, Loop
Expansions and the Cosmological Constant,'' Nucl. Phys. {\bf B357} (1991) 534.}
\lref\duffss{M. J. Duff,  ``Strong/Weak Coupling Duality from the Dual
String,'' NI-94-033,
CTP-TAMU-49/94, hep-th/9501030.}
\lref\trrev{For a review see  C. G. Callan, J. A. Harvey and A. Strominger,
``Supersymmetric String
Solitons'' in String Theory and Quantum Gravity 1991: Proceedings of the
Trieste Spring School,
World Scientific, Singapore, 1991, hep-th/9112030.}
\lref\ch{C. Callan and J. A. Harvey, Nucl. Phys. {\bf B250} (1985) 427.}
\lref\SCH{J. H. Schwarz, Nucl. Phys. {\bf B226} (1983) 269. }
%
%

\Title{\vbox{\baselineskip12pt
\hbox{EFI-95-16}
\hbox{hep-th/9504047}}}
{\vbox{\centerline{\bf{THE HETEROTIC STRING IS A SOLITON} }}}
{
\baselineskip=12pt
\centerline{Jeffrey A. Harvey}
\bigskip
\centerline{\sl Enrico Fermi Institute, University of Chicago}
\centerline{\sl 5640 Ellis Avenue, Chicago, IL 60637 }
\centerline{\it Internet: harvey@poincare.uchicago.edu}
\bigskip
\centerline{ Andrew Strominger }
\bigskip
\centerline{\sl Department of Physics}
\centerline{\sl University of California}
\centerline{\sl Santa Barbara, CA 93206-9530}
\centerline{ \it Internet: andy@denali.physics.ucsb.edu }

\bigskip
\centerline{\bf Abstract}
It is shown that the Type IIA superstring
compactified on $K3$ has a smooth string soliton with the same zero mode
structure as  the heterotic string compactified on a four torus, thus
providing new evidence for a conjectured exact duality between the two
six-dimensional string theories. The chiral worldsheet bosons
arise as zero modes of
Ramond-Ramond fields of the IIA string theory and live on a
signature $(20,4)$ even, self-dual  lattice.  Stable, finite loops of
soliton string provide the charged Ramond-Ramond states necessary
for enhanced gauge symmetries at degeneration points of
the $K3$ surface.
It is also shown
that Type IIB strings toroidally compactified to six dimensions have a
multiplet of string solutions with  Type II worldsheets.
}
\Date{April, 1995}
%

\newsec{Introduction}

The idea that fundamental strings might be viewed as solitons is an old one
dating back at least to the work of Nielsen and Olesen \no.
While past attempts to describe fundamental strings as solitons
in an ordinary field theory have
not been fruitful, from the current
perspective one might hope to find them
as solitons in another dual string theory.
Some weak evidence for such a possibility
was presented in \refs{\DAHA, \DGHR} where certain analogies between
fundamental
strings and solitons were drawn.

However duality in the geometrical sense does not usually relate a fundamental
string
to a soliton string. A simple generalization \tei\ of Dirac's argument shows
that, in $d$
spacetime dimensions, a $p$-brane\foot{We adopt the
standard terminology that a $p$-brane is an extended object with $p$ spatial
dimensions.} is dual to a $(d-p-4)$-brane (Thus in ten dimensions a fivebrane
is dual to a string and should couple to the dual form of
supergravity \duffive.). Unobservability of the Dirac $(d-p-3)$-brane emanating
from the
$p$-brane ({\it i.e. } the generalized Dirac string) implies that the
minimal charge carried by the dual $(d-p-4)$-brane is inversely proportional to
the charge carried by the $p$-brane. Thus the $(d-p-4)$-brane will be weakly
coupled
when the $p$-brane is strongly coupled and vice versa. Motivated in part by
this observation it was conjectured in \hetsol\
that fivebrane solitons should
be relevant for the description of the strongly coupled
phase of fundamental ten-dimensional string (onebrane) theory. This is the
natural generalization to ten-dimensional string theory of the
Montonen-Olive conjecture for $d=4$ Yang-Mills theory.  However
as very little is understood about the quantum theory of fivebranes
(see \townrev\ for a review)
this conjecture is
difficult to test.

In six dimensions, a string is dual to a string and the prospects for a
duality which can be understood concretely
are much more promising \refs{\duffss, \wit}. In addition, as discussed
in \refs{\schsen,\duffss,\wit}\foot{Reference \schsen\ derived these results
using a toroidally compactified version of string-fivebrane duality which
is equivalent to string-string duality in six dimensions.},
string-string duality in six dimensions  can be used to deduce
$S$-duality
in four-dimensional toroidally compactified heterotic string theory, because it
transforms
$S$-duality into the perturbatively manifest $T$-duality of the dual theory.
This is referred to as the ``duality of dualities''.
Evidence for $S$-duality of the toroidally compactified heterotic string has
been accumulating
\sdualref\ and there have been a variety of tantalizing
hints of connections between
this $S$-duality and dualities in various dimensions involving strings and/or
higher $p$-branes \refs{\drev,\town}.   In \wit\ a
coherent picture was presented of how these
various dualities should be related to each other. One important lesson
is that a consistent picture of strong coupling string dynamics in dimensions
four through seven emerges as a consequence of a six-dimensional duality
between the  Type IIA string on $K3$ and the heterotic string on $T^4$.

In this paper we shall
investigate the worldsheet action for string solitons arising in several
six-dimensional string theories. These string solitons are simply the fivebrane
solitons of \refs{\hetsol,\duffone, \wsheet, \wbran}
with four of their dimensions wrapped around the internal four-space
used in the compactification from ten to four dimensions,
and correspond to exact conformal field theories \wsheet.

We first consider the nonchiral IIA theory on $K3$. Assuming a
quantization condition on
Ramond-Ramond charge, we will show that the
world sheet theory of the string
soliton has a chiral structure with right-moving supersymmetry. In addition to
the four transverse bosonic zero modes and eight right-moving
fermion zero modes there are 20 left-moving and 4 right-moving bosonic zero
modes which lie on an even, self-dual Lorentzian lattice with signature
$(20,4)$. This is precisely the structure of the
heterotic string
with a generic $T^4$ compactification! The soliton string couples to the dual
Kalb-Ramond field and
has string tension proportional to $ e^{-2\phi }/\apm$
so that the fundamental string loop coupling constant
becomes the dual worldsheet sigma-model
expansion parameter\refs{\hetsol, \duffcq, \wbran}.
We will also identify closed string solitons which become massless and lead to
enhanced gauge symmetries at special points in the $K3$ moduli space,
in accord with conjectures in \refs{\town,\ferr,\wit}.

The emergence of the heterotic string as a soliton is
quite remarkable. While the existence of a rank-three
field strength more or less implies the existence of string solitons in six
dimensions, there is no guarantee that the effective worldsheet dynamics
describe a critical string theory, let alone the
intricate chiral structure of the heterotic string.
It is especially
striking that such a structure  arises from the non-chiral
IIA string.

Upon further toroidal compactification to four dimensions, the
``duality of dualities'' implies that the
$T$-duals of the BPS states \DAHA\ of our type II string soliton are the
$S$-duals
of the BPS states of the toroidally compactified heterotic string.
Thus in principle our results provide the solution
to the $H$-monopole problem, but it remains to be understood
how this works in detail. We hope to report on this elsewhere.

We will also show  that
the worldsheet action for the chiral type IIB string compactified on $T^4$
is a type II string with a $T^4$  target space as are the
string solitons of the IIA theory.

These results provide evidence in favor of (and were in part motivated by)
the recent conjectures that the IIB theory is dual to itself
while the IIA theory on $K3$ is dual to the heterotic string on $T^4$
\refs{\town, \wit, \vafa}. The IIA-heterotic duality conjectures were
based largely on the the earlier observation \nati\ that equivalence of
the low-energy effective theories follows from symmetry considerations
and on the consistent picture of dynamics in other dimensions that
it implies.
The emergence of the heterotic string
as a soliton requires several ``miracles'' that do not follow from symmetry
considerations
alone and thus provides substantial new evidence for exact string-string
duality.
We will also find similar evidence for the conjecture \refs{\town,\towne,\wit}
that
eleven-dimensional supergravity is dual to the heterotic string
compactified on $T^3$.

While  these results are intriguing, a number of puzzles remain.
One of these concerns  toroidally compactified heterotic strings.
The conjectured six-dimensional  IIA - heterotic duality suggests that the
heterotic string should have a
string soliton  with a worldsheet which is that of a IIA theory with a $K3$
target. This
structure does not appear to emerge in any simple way. Indeed,
as pointed out in \duffss, toroidal
compactification of the heterotic fivebrane seems to lead to a
non-critical string\foot{Presumably of the general variety
discussed in \jpas.}. This puzzle is  related to the $H$-monopole problem
\gau\ encountered in attempts to verify four-dimensional
$S$-duality, since the $H$-monopoles arise in compactification of the
six-dimensional string solitons.  It seems likely that this puzzle can be
addressed by a generalization of the techniques discussed here, at least
at special points in the $K3$ moduli space.

A second puzzle is the relation of this
string-string duality to the string-fivebrane duality of \hetsol. The
fact that the string solitons are compactified fivebranes clearly  suggests a
connection: string-string duality may be viewed as a compactified,
six-dimensional
version of string-fivebrane duality. However while
in six dimensions it is sensible to
describe the strongly coupled dynamics by a perturbative quantization which
treats the weakly-coupled string solitons as
fundamental objects, it is not clear if an analogous
description is
sensible in ten dimensions since we do not know how to
quantize fundamental fivebranes. The proper role
of fivebranes in nonperturbative string dynamics thus remains a mystery.

Finally, it is worth pointing out that the proposed IIA- heterotic string
duality
takes the smooth soliton solution described herein into a singular solution
which
describes the field around a fundamental heterotic string \DGHR. It has been
appreciated for some time that the question of the singularity of the
fundamental
string solution depends on the metric used and that in the dual, or
``fivebrane'' metric the solution  is in fact non-singular \refs{\wbran,
\drev}, in harmony
with the idea that fundamental strings are solitons of a dual theory.
It has also been appreciated for some time \senone, that properties of
the low-energy effective theory imply that the low-energy
fields outside a heterotic string could carry long range fields
corresponding to the possibility of charge excitations on the string. The key
missing
ingredient in these previous observations was identification of the source of
these
long range fields.  In the present work we fill this gap
by showing that the full chiral zero mode structure of the fundamental
heterotic string arises directly in the dual theory through well defined,
normalizable zero modes of a non-singular soliton solution. With this in hand
one
can quantize the solitons as elementary objects and verify that their spectrum
is precisely that required for an exact string-string duality.

The organization of this paper is as follows.
Section 2 contains the construction of string solitons
in the Type IIA theory on $T^4$ and $K3$.  We show that the heterotic string
arises as a soliton in the $K3$ case and relate the zero mode structure to
the intersection matrix of $K3$ and quantization of Ramond-Ramond (RR) charge.
We also argue that the quantum of RR
charge must have a specific value.  Section 3 summarizes an analogous
construction in eleven-dimensional supergravity compactified on $K3$.
Section 4 extends these results to the Type IIB theory on $T^4$ by constructing
string solitons with Type II worldsheets and in Section 5 we mention an
extension of these solutions which follows from an $SL(5,Z)$ duality of
the Type IIB theory on $T^4$. In Section 6 we summarize our results and
discuss some of their implications.  We point out that finite loops of soliton
string carry RR charge and hence must be stable and  at special
points in the $K3$ moduli space can provide  the massless states with
the required degeneracies and quantum numbers needed for
enhanced gauge symmetry.

\newsec{Heterotic String as a Soliton in Type IIA String Theory}

In ten dimensions the
IIA string theory\TEN\
contains a two-form field strength $G$ and a four-form field strength $F$
as well as the three-form $H$. The bosonic part of the low-energy action is
(in sigma-model variables)
\eqn\iia{\eqalign{{S} &={1 \over 16 \pi {\apm}^4 }
{\int} {d^{10}}{x}{\sqrt{-g}}\biggl[ {e^{-2\phi}}({R+4}
({\nabla}{\phi})^2
-{1\over3}{H^2})-\apm {G^2}-{\apm \over12}{F^{\prime 2}}\cr
&~~~~~~\qquad \qquad  \qquad \qquad
-{\apm \over 288}\epsilon^{M_1...M_{10}}F_{M_1M_2M_3M_4}F_{M_5M_6M_7M_8}
B_{M_9M_{10}}\biggr]\cr
&={1 \over 16 \pi {\apm}^4 }
{\int} {d^{10}}{x}{\sqrt{-g}} {e^{-2\phi}}({R+4}
({\nabla}{\phi})^2)\cr
&~~~~~ -{1 \over 8 \pi{\apm}^4} \int(e^{-2\phi}H\wedge *H+\apm G\wedge *G+\apm
F^{\prime}
\wedge * F^{\prime}+2 \apm F\wedge F\wedge B) .}}
where $G=dA,~~H=dB$,  $F=dC$ and $F^{\prime}=dC+2A\wedge H$.\foot{In
components $G_{MN}=2\partial_{[M}A_{N]},~~H_{MNP}=3\partial_{[M}B_{NP]},~~
F^{\prime}_{MNPQ}=4\partial_{[M}C_{NPQ]}+8A_{[M}H_{NPQ]}$.}

The spacetime of the string soliton  has the form
\eqn\decomp{K \times R^4 \times M^{1,1},}
where $R^4$ is the space transverse to the string and $M^{1,1}$ is the string
world-sheet.
The string vacuum equations of motion require that $K$ be a Ricci-flat
hyperkahler manifold, {\it i.e.} $R^4$, the four-torus $T^4$ or a $K3$ surface.
We will
consider the latter two possibilities ($R^4$ leads back to fivebranes).
When necessary we will denote the coordinates on the spacetime \decomp\  by
$X^M = (y^a, x^\mu, \sigma^\alpha)$.
The string soliton solution is given by choosing a Ricci-flat hyperkahler
metric
on $K$, a flat metric on $M^{1,1}$, and the following field configuration on
the
transverse space:
\eqn\symm{\eqalign{ {e^{2\phi}} &= {e^{2\phi_0}}  +
                                      {   \alpha^{\prime} \over x^2} ,\cr
{H}_{\mu\nu\lambda} &= {-}{\epsilon_{\mu\nu\lambda}}^{\rho}{\nabla}
_{\rho}{\phi},\cr
{g_{\mu\nu}} &= {e^{2\phi}}{\delta_{\mu\nu}}.\cr}  }
with all other fields vanishing.  This soliton string carries the minimal
 ``magnetic'' charge
\eqn\hch{ Q = - {1 \over 2 \pi^2} \int_{S^3} H = \alpha', }
in contrast to the ``electric'' charge carried by  the fundamental Type IIA
string.
Note that the solution is presented  in coordinates in which the metric is
asymptotic to $e^{2\phi_0}\delta_{\mu\nu}$
rather than $\delta_{\mu\nu}$.  This solution is just the ``symmetric
fivebrane''
solution of \refs{\wsheet, \wbran}
(closely related solutions are discussed in
\refs{\sswt, \hetsol, \rey, \duffone} ) reinterpreted as a
string soliton in six
dimensions, and as such much of the following discussion parallels that of
\wbran.
The solution \symm\ is a gravitational analog of a Yang-Mills instanton
with the generalized connection given by the sum of the
spin connection and the torsion,
\eqn\gencon{ {\Omega_{+ \mu}}^{ij} = {\omega_\mu}^{ij} + {H_\mu }^{ij}  , }
being an anti-self-dual $SU(2)$ connection.  The solution \symm\
is known to correspond to an exact  $(4,4)$ superconformal field theory
\refs{\wsheet,\trrev}
and thus is an exact classical solution of the IIA string theory.

The
supersymmetry transformation laws in the soliton background are
\eqn\twoasol{\eqalign{\delta\psi_{\pm M}&=
          \nabla_M\epsilon_\pm\mp{1\over4}H_{MNP}
  \Gamma^{NP}\epsilon_\pm ,\cr
\delta\lambda_\mp&=\Gamma^M\nabla_M\phi\epsilon_\pm\pm{1\over6}H_{MNP}
  \Gamma^{MNP}\epsilon_\pm ,\cr }}
\noindent where $\psi_{\pm M}$ and $\lambda_\mp$ are respectively the
gravitinos
and dilatinos and the subscript $\pm$ denotes the ten-dimensional chirality.
Decomposing the ten-dimensional Lorentz group in accord with \decomp\
we have $SO(9,1) \rightarrow SO(4) \times SO(4) \times SO(1,1) $
and
\eqn\epde{\eqalign{ \epsilon_+ & \rightarrow (2_+, 2_+)^+ + (2_+,2_-)^-
                        + (2_-,2_+)^- +  (2_-,2_-)^+ ,\cr
                        \epsilon_- & \rightarrow (2_+,2_+)^- +
(2_+,2_-)^+           + (2_-,2_+)^+ + (2_-,2_-)^- ,\cr
}}
where subscripts refer to the the $SO(4)$ chirality and the superscript
indicates
the $SO(1,1)$ chirality.

The vacuum configuration on $T^4$ preserves all
of the supersymmetries \epde\  and leads to a six-dimensional theory with
$(2,2)$ six-dimensional supersymmetry. The soliton string configuration \symm\
breaks
the $\epsilon_+$ supersymmetries which transform as $2_+$ under the second
$SO(4)$ factor
and the $\epsilon_-$ supersymmetries which transform as $2_-$.  Thus on $T^4$
the completely unbroken supersymmetries are
\eqn\unbroke{  (2_+,2_-)^- + (2_-,2_-)^+ + (2_+,2_+)^- + (2_-,2_+)^+  , }
and the supersymmetries broken by the string soliton are
\eqn\strbroke{ (2_+,2_+)^+ + (2_-,2_+)^- + (2_+,2_-)^+ + (2_-,2_-)^-  .}
The latter lead to fermion zero modes on the string soliton.

The vacuum configuration on $K3$  preserves the supersymmetries $(2_+,2_+)^+ +
(2_+,2_-)^- + (2_+,2_+)^- + (2_+,2_-)^+ $ and leads to a theory with
$(1,1)$ six-dimensional supersymmetry.
The string soliton background  breaks an additional half of the
supersymmetries.
The completely unbroken supersymmetries are
\eqn\unbrokeb{ (2_+,2_-)^- + (2_+,2_+)^-  ,}
while the supersymmetries unbroken by $K3$ but broken by the
soliton string background are
\eqn\strbrokeb{ (2_+,2_+)^+ + (2_+,2_-)^+ .}
These broken supersymmetries lead
to 8 real  supertranslation zero modes on the string \hupo\
which are clearly chiral
on the world sheet.  We will refer to these as right-moving fermion zero modes.
Thus the non-chiral IIA string
theory when compactified on $K3$ leads to a chiral static worldsheet action for
the string soliton.

The most obvious bosonic zero modes of the soliton string are the four
translation
zero modes associated to translation in the four dimensions transverse
to the string. Introducing four collective coordinates $X^\mu (\sigma)$ for
these fields one finds the effective action\foot{See references \refs{\DGHR,
\hetsol, \wbran} for a more complete discussion of this type of effective
action.}
\eqn\wsact{ {N \over {\apm}}
\int d^2\sigma e^{-2\phi}( \eta^{\alpha \beta}\partial_\alpha
X^\mu\partial_\beta X^\nu
g_{\mu\nu}+ \epsilon^{\alpha\beta}\partial_\alpha X^\mu\partial_\beta X^\nu
\tilde B_{\mu\nu}),}
where $N$ is a normalization factor (proportional to $V_K / {\alpha'}^2 $ with
$V_K$ the
volume of $K$) and $\tilde B$
is the dual Kalb-Ramond field defined by $e^{-2\phi}*H=d\tilde B$ in
six dimensions.
This is a static gauge ($X^\alpha=\sigma^\alpha$) form of the
Polyakov or Nambu action. As argued in \hetsol\ (in the context
of fivebranes) the powers of
$e^{-2\phi}$ are simply determined by scaling properties of the solutions
under constant shifts of $\phi$, and the coupling to $\tilde B$ follows from
the required violation of the Bianchi identity for $ H$ in the presence of the
soliton string.  The former statement has the immediate interesting
consequence,
as emphasized in \duffcq, that
quantum corrections on the string soliton worldsheet are
controlled by the loop expansion parameter of the
original string theory.

The {\it unbroken} spacetime supersymmetries imply worldsheet
supersymmetry \hupo, which in turn requires the existence of four
additional bosonic zero modes to pair with the right-moving fermion
zero modes.  Worldsheet supersymmetry makes no obvious prediction
for left-moving bosonic zero modes. Extra bosonic zero modes arise
by considering the equation for  small fluctuations of the three-form potential
$C$ in the soliton
background. In ten-dimensional form this equation is
\eqn\smallfluc{ d
 {\ast} F' = - 2 F \wedge H .}
%
%
To solve this we write an ansatz\foot{This corrects equation 5.10 of \wbran.}
\eqn\ansatz{ C =  {\apm \over 2 \pi} X^I(\sigma) U_I(y) \wedge
d e^{2\phi_0-2 \phi(x)} ,}
with $X^I$ a worldsheet zero mode,
$e^{2 \phi(x)}$ the
background dilaton field of the string soliton and $I= 1,...b_2$ with
$b_2$ the
second Betti number of $K$. The harmonic two-forms
$U_I$ comprise
an integral basis for $H^2(K,Z)$. Note that \ansatz\ is normalizable and
localized
near the ``throat'' of the string soliton.  The equation of motion \smallfluc\
then requires that $X^I$ be a free two-dimensional field and,
using $H=-{\hat \ast}d\phi$ and $dH=0$,
that
\eqn\wdual{  d X^I \wedge U_I = {\hat \ast}d X^I \wedge {\hat \ast} U_I .}
We use ${ \ast}$ for the ten-dimensional Hodge dual and
wherever ${\hat \ast} $ appears acting on a form tangent to one component of
the
decomposition \decomp\ it acts as the Hodge dual within that component.

Since the Hodge dual of a harmonic form is harmonic, the duals of the $U_I$ may
be expressed as a linear combination
\eqn\hdl{ {\hat \ast} U_I=U_J{H^J}_I,}
where $ {H^J}_I$ depends on the moduli of $K$.
This enables us to rewrite the constraint \wdual\ as
\eqn\rwr{ \partial_\pm X^I=\pm{H^I}_J\partial_\pm X^J,}
where $\partial_\pm$ are left and right
worldsheet derivatives. The fact that $\hat \ast \hat \ast =1$ implies
\eqn\hsq{{H^I}_J{H^J}_K={\delta^I}_K,}
so that $H$ has $b_2^\pm$  eigenvalues  $\pm 1$, where $b_2^\pm$  is
the number of self-dual  (anti-self-dual) two forms on $K$.

An action for the $X^I$ can be obtained by dimensional reduction of the ten
dimensional action \iia .
After integrating the zero mode wave functions \ansatz\ over $K \times R^4$
one obtains
\eqn\xmet{S_{X^I}={1 \over 4 \pi}\int d^2\sigma L_{IJ}\partial_+ X^I\partial_-
X^J ,}
where
\eqn\kint{\eqalign{L_{IJ}&\equiv \int_{K}U_I\wedge U_J \cr
}}
is the intersection matrix on $H^2(K,Z)$.
This action is of course subject to the chiral constraint \rwr.
We further note that $\hat \ast \hat \ast =1$ implies
\eqn\whsq{{H^K}_I L_{JK}=L_{IK} {H^K}_J .}

We are still short one bosonic zero mode needed for worldsheet supersymmetry.
The missing zero mode involves
a combination of $C$ and the
one-form potential $A$ whose field strength $G$ obeys the equation of motion
\eqn\geqn{ d { \ast} G= -2 { \ast} F'  \wedge H  .}
A solution of \geqn\ and \smallfluc\ is given by
\eqn\azero{\eqalign{A&=  {1 \over 2 \pi} X^0(\sigma) de^{2 \phi_0-2\phi},\cr
C&=-{1 \over \pi} X^0(\sigma) e^{2 \phi_0 -2\phi} H,\cr}}
provided that $X(\sigma)$ is a free two-dimensional field.  This thus gives
an additional $(1,1)$ (left,right)-moving bosonic zero mode.
Note that since under
abelian gauge transformations $\delta A=d\epsilon$ and $\delta C=-2\epsilon H$,
this zero mode is equivalent under a `large' gauge transformation to a
configuration with vanishing $C$. $X^0$ is governed by the action
\eqn\xmt{S_{X^0}={V_K \over 2 \pi {\apm}^2}\int d^2\sigma \partial_+
X^0\partial_- X^0  .}

Naively the worldsheet fields $X^I$ have a  noncompact range,
and the solitons have a continuous spectrum and pathological thermodynamics.
However we believe the $X^I$s
should be periodically identified for the following reason.
In ten dimensions there exists a conserved ``electric'' charge
\eqn\fquant{q=\int_{\Sigma^6}*F,}
and its  ``magnetic''
dual
\eqn\fmquant{\tilde q=\int_{\Sigma^4}F.}
Elementary string states do not carry this charge, but the electric (magnetic)
charge is
carried by black twobrane (fourbrane)
 solutions of the theory\bhole\ (as well as by momentum/winding states of
fivebranes).
If both types of objects are included in the
theory, the arguments of \tei\ then require that the product $q\tilde q$
is quantized. We therefore assume a quantization condition
\eqn\fqnt{\int_{\Sigma^4}F=ng_4\apm,~~~n \in Z,}
for some dimensionless minimal magnetic charge $g_4$. $g_4$ does not enter
in to any perturbative
calculation. Later we shall fix $g_4$ by nonperturbative considerations.

As in ordinary electromagnetism, if the charges of all states are
quantized, gauge transformations
are periodic.  The theory is invariant under two-form gauge transformations
\eqn\trg{ \delta C = d\epsilon_2.}
Under such a transformation a state corresponding to a minimal
electrically-charged
twobrane
with a spatial surface $\Sigma_2$ acquires a phase
\eqn\twphs{ \exp \left( {2 \pi i \over g_4\apm}\int_{\Sigma_2}\epsilon_2
\right) .}
If $\epsilon_2$ is ${ g_4} \alpha'$ times an element of $H^2(K,Z)$
this will be unity for all twobranes. We should therefore identify
\eqn\eid{\epsilon_2\sim\epsilon_2+{ g_4}\apm n^I U_I,}
for arbitrary integers $n^I$.

For {\it constant} $X^I$, the zero mode \ansatz\ is a gauge transformation
with parameter
\eqn\prmt{\epsilon_2={\apm \over 2 \pi}X^IU_I}
far from the soliton. The identification \eid\ therefore
implies the key identification
\eqn\xid{X^I \sim X^I+{2 \pi  g_4} n^I .}

The conditions
on the $X^I$s may look a little more familiar in terms of the rotated
fields
\eqn\yxo{Y^J={O^J}_IX^I,}
where $O$ is an element of $SO(b_2^+,b_2^-)$ obeying
\eqn\ostf{\eqalign{{O^I}_K {H^K}_L {(O^{-1})^L}_J&={\eta^I}_J ,\cr}}
and $\eta$ is diagonal with $ b_2^+$ plus ones and $ b_2^-$  minus ones.
The $Y^I$s are pure left or pure right-moving fields,
\eqn\ylr{\partial_\pm Y^I=\pm {\eta^I}_J\partial_\pm Y^J,}
with metric $ OLO^T$.
The identifications \xid\ become
\eqn\yid{Y^I \sim Y^I+ {O^I}_Jn^J2\pi g_4 .}
This identification defines a lattice whose
vectors are $Ong_4$.
The inner product of two lattice vectors
is
\eqn\lvct{\eqalign{{O^I}_Jn^J [{(O^{-1})^M}_IL_{MN}{(O^{-1})^N}_K ]{O^K}_L
m^Lg_4^2
&=n^IL_{IJ}m^J g_4^2.\cr
}}
Properties of the intersection matrix $L$ on
$H^2(K,Z)$ then imply that this is an even self-dual Lorentzian lattice
if and only if
\eqn\gqnt{g_4=1.}
We henceforth assume this to be the case.

Similar considerations suggest a quantization condition
\eqn\twqnt{\int_{\Sigma^2}G=ng_2,~~~n \in Z.}
As in the preceding this leads to the periodic
identification
\eqn\xzid{X^0\sim X^0+{2 \pi  g_2}.}

When the Kalb-Ramond field $B$ is a nontrivial element of $H^2$ on $K$, the
$\int F\wedge F \wedge B $ term in \iia\ leads to a
shift in the quantization condition.
We have not worked this out in detail, but the modified
quantization condition will mix up $X^0$ and $X^I$, and will lead to
dependence of the worldsheet theory on the moduli corresponding to $B$.

When the theory is reduced to six dimensions on $K$ there are $2+ b_2(K)$
$U(1)$ gauge fields in the low-energy theory which come from Ramond-Ramond
fields.  One of these comes from the original $U(1)$ field strength $G$. A
second
arises from the dual field strength ${\hat \ast} F$ which is a two-form
in six dimensions. The other $b_2(K)$ gauge fields
arise from writing $C = U_I \wedge A^I$ with
$A^I$ one-form gauge potentials in six dimensions.
Since the $X^I$ at zero-momentum are pure gauge the periodicity
of the $X^I$ is equivalent to saying that the six-dimensional $U(1)$ gauge
groups
are compact.

We can now summarize our results for the two choices $K=K3, ~~T^4$.

\medskip
{\bf K3}
\medskip
For $K3$, $b_2^+(K3)= 3$, $b_2^-(K3) =19$ and
\eqn\kper{L_{IJ}=\left[\Gamma_8 \oplus \Gamma_8 \oplus \sigma^1\oplus
\sigma^1\oplus \sigma^1\right]_{IJ} , \qquad \sigma^1 = \pmatrix{
0 & 1 \cr
1 & 0 \cr
}}
where $\Gamma_8$ is the Cartan matrix for $E_8$.  The zero modes
\ansatz\ and \azero\ therefore give
$(20,4)$ (left,right)-moving
bosonic zero modes  with periodicities given by \xid\
and \xzid. It follows from \rwr\ that these zero modes live
on an even self-dual $(20,4)$ Lorentzian lattice.
There are also 8 right-moving worldsheet fermions as required by the
unbroken supersymmetries, which transform
in the $(2_+,2_+)^+ + (2_+,2_-)^+$ of the spacetime
$SO(4)\times SO(4)\times SO(1,1)$. Together with the transverse bosonic
coordinates
of the string, this gives the Green-Schwarz, static gauge
worldsheet theory of a Narain compactification to six dimensions of the
heterotic string.
So, we have finally learned ``why'' the $K3$ intersection matrix and the
heterotic string both involve $E_8 \times E_8$!

The moduli
space of Ricci flat metrics on $K3$ is
\eqn\modn{ { \cal N }= SO(19,3;Z)\backslash
SO(19,3;R)/SO(19;R) \times SO(3;R) . }
As the $K3$ space is varied, the condition
\rwr\ varies, and the dual
soliton varies over a 58-dimensional subspace of Narain compactifications.
Inclusion of the effects of $B$ introduces 22 additional moduli. It is known
\nati\
that these moduli complete the space \modn\ to
\eqn\modm{ {\cal M} = SO(20,4;Z) \backslash
SO(20,4;R)/SO(20;R) \times SO(4;R) }
Although we
have not worked out the details of the effects of the $B$ moduli on the
zero mode quantization, one presumably  obtains the full space \modm\
from the collective coordinate expansion of the heterotic soliton
string, now viewed as the moduli space of Narain compactifications.

Our collective coordinate expansion provides a precise map from the moduli
space of  $K3$ compactifications of the IIA string to $T^4$
compactifications of the heterotic string. It will be interesting to study the
precise structure of this map for various degenerations of the $K3$
moduli space.

\medskip
{$\bf T^4$}
\medskip
For $T^4$, $b_2^+=b_2^-=3$ and
\eqn\tper{L_{IJ}=\left[ \sigma^1\oplus \sigma^1\oplus \sigma^1\right]_{IJ}
.}$X^I$ and $X^0$ therefore parameterize a  self-dual
$(4,4)$ Lorentzian lattice. This is related to the lattice defining the
original
$T^4$ in a non-trivial but calculable manner. In this case the compactification
does not break any supersymmetries, and the supersymmetries
which are unbroken by the soliton imply both left and right
worldsheet supersymmetry. Invoking the equivalence \refs{\polc, \dhs} (under
$T$-duality) of toroidally
compactified IIA and IIB strings, this soliton string may be identified
as either a
IIA or a  IIB string with different target tori.

\newsec{Eleven-Dimensional Supergravity}
It was conjectured in \refs{\town,\towne} that there should be an exact
duality of  eleven-dimensional supergravity, defined as a fundamental
supermembrane
theory, with the ten-dimensional type IIA theory. In \wit\ evidence
for such a relation was given at the level of their low-energy
effective field theories, where one can avoid the sticky
issue of quantization of supermembranes. Together with
other observations, this would imply
a duality relating $d=11$ supergravity compactified on
$K3$ to the heterotic string compactified on $T^3$. Evidence for this
is provided by the fact that
$d=11$ supergravity compactified on
$K3$ has string solitons whose worldsheet dynamics are precisely those
of the heterotic string compactified on $T^3$. This follow easily from the
results of the previous section. Since the low-energy effective IIA theory
was originally obtained \TEN\ by compactification of  $d=11$ supergravity,
the equations which must be solved are quite similar. The $d=11$
theory contains a fivebrane solution \refs{\guven,\wbran} which becomes a
string
when four of its dimensions wrap around the $K3$.
One finds all the zero modes
\ansatz\ in a nearly identical construction. The extra zero mode \azero\ is
absent
however because the corresponding fields arise only after reduction to $d=10$.
Thus one obtains the desired heterotic soliton string with bosonic fields
living on
a $(19,3)$ Lorentzian  lattice. Perhaps, following the lead taken in \towne,
one may view
$d=11$ supergravity on $K3$ as a consistent quantum
theory whose perturbative quantization
is defined
by a strong-coupling expansion of quantized string solitons rather than as a
weak coupling expansion of the original fields!

\newsec{Type IIB String Duality}

We now turn to the chiral type IIB theory \SCH.
In ten dimensions,
the IIB theory involves a five-form field strength $F$ which is self-dual.
Because of this there is no simple covariant action for the IIB theory.
The field content, equations of motion,
and on-shell supersymmetry transformation
laws are known and are sufficient to discuss the general properties
of the solution. The bosonic fields that enter into the fermion
transformation laws are (in unitary gauge \SCH ),
in addition to the metric and $F$, a
{\it complex} closed three-form
field strength $H=dB$ and a {\it complex} scalar ${\cal B}$ obeying $|{\cal
B}|<1$.
The ten-dimensional
supersymmetry transformation
laws of the fermion fields in the soliton background  are given by
\eqn\twobsusy{\eqalign{
\delta\lambda&={i}\Gamma^MP_M\epsilon^*-{i\over24}\Gamma^{MNP}
G_{MNP}\epsilon ,\cr
\delta\psi_M&=D_M\epsilon+ {i \over 480} \Gamma^{PQRST} F_{PQRST}
\Gamma_M \epsilon +
{1\over96}(\Gamma_M\/^{NPQ}G_{NPQ}
-9\Gamma^{NP}G_{MNP})\epsilon^* ,\cr }}
where
\eqn\pgdef{\eqalign{P_M&={\partial_M {\cal B} \over 1-{\cal B}^*{\cal B}},\cr
G_{MNP}&={H_{MNP}-{\cal B}H^*_{MNP} \over \sqrt{1-{\cal B}^*{\cal B}}},\cr}}
in the notation of  \refs{\SCH,\GSW} (with $\kappa=1$
and $\Gamma_{11}\epsilon=-\epsilon$).
We consider this theory in the spacetime
\eqn\decomp{T^4 \times R^4 \times M^{1,1}.}
(Soliton strings with $K3$ compactification are apparently non-critical
strings.)
For the string soliton,
the non-zero components of the fields are
\eqn\twobbrane{\eqalign{P_\mu&={1\over 2}\nabla_\mu\phi,\cr
           G_{\mu\nu\rho}&=-2\epsilon_{\mu\nu\rho}
           \/^\lambda\nabla_\lambda\phi,\cr
{\cal B}&={\rm tanh}{\phi \over 2},\cr
e^{2\phi}&=e^{2\phi_0}+{\alpha^\prime\over x^2},\cr
g_{\mu\nu}&=e^{3\phi/2}\ \delta_{\mu\nu},\cr
g_{\alpha\beta}&=e^{- \phi/2}\eta_{\alpha\beta}.}}
The metric which appears in these equations is the ``standard'' general
relativity
metric and is related to the string sigma model metric by a factor of
$e^{-\phi/2}$.

For this field configuration, $\delta\psi$ and $\delta\lambda$ vanish for 16
choices of the spinor $\epsilon$.  The other 16 generate 16 fermion zero modes.
Worldsheet supersymmetry implies that in static gauge the number of fermi
fields must be twice the number of bose fields. There are the usual
four bosonic translation zero modes, but  we are still short four
bosonic zero modes.  The four extra zero modes are contained in the excitations
of antisymmetric tensor fields with
%
%
the zero mode field strengths  given by\foot{This corrects formula 5.6 of
\wbran.}
\eqn\zermodes{\eqalign{F_0&=e^{-\phi/2}dX^I_\pm(\sigma)(V_I\wedge G \mp
2W_I\wedge d\phi),\cr
                       G_0&=4ie^{-\phi/2}dX^I_\pm(\sigma)V_Id\phi .}}
\noindent where $V_I$ ($W_I$) comprises a
harmonic basis for $H^1(T^4,Z)$ ($H^3(T^4,Z)$).  For pure right or
left-moving excitations the
ansatz \zermodes\ preserves half of the supersymmetry preserved by
the solution \twobbrane\ and
it is easy to check that $F_0$ is
a self-dual five-form provided
\eqn\dxx{dX^I_\pm=\pm {H^I}_J{\hat \ast}dX^J_\pm,}
where ${H^I}_J$ relates the Hodge-dual and Poincare-dual bases
\eqn\hexp{{H^K}_I W_K=\hat \ast V_I.}

As before a quantization
conditon on the RR charges will lead to periodic identifications of the
zero modes.
Equation \dxx\ then gives four left-moving and four right-moving worldsheet
bosons on a $(4,4)$ self-dual Lorentzian lattice. Unbroken worldsheet
supersymmetries then relate these to both left and right-moving superpartners,
and one obtains a type II worldsheet.

\newsec{Extended Duality in Type II String Theories }

This is not the end of the story for the IIB string. In fact there are
more string solutions in six dimensions.
To see this let us first go back to
ten dimensions, in which there is a fundamental string solution
\refs{\DAHA,\DGHR}
with a singularity at the origin corresponding to an
macroscopic fundamental string as a source.

The equations describing the low-energy effective IIB dynamics have an
$SU(1,1)$ symmetry. As noted in \wbran, charge quantization clearly forbids a
continuous symmetry, but allows a discrete $SL(2,Z)$ subgroup
which was conjectured in \town\ to be an exact symmetry of the ten-dimensional
theory. Under this $SL(2,Z)$ the real and imaginary
parts $H_1$ and $H_2$ of $H$ are a doublet
\eqn\ztwo{\eqalign{(H_1,H_2)&\rightarrow (aH_1+bH_2, cH_1+dH_2), \cr}}
with $a,b,c,d\in Z$ and $ad-bc=1$. Since everything transforms covariantly
under these transformations one obtains new string
solitons.

The usual fundamental string of \refs{\DAHA,\DGHR} acts as a source for the
usual
NS-NS Kalb-Ramond field $H_1$. Under the $SL(2,Z)$ transformation
$H_1\rightarrow H_2$ this transforms into
a string which acts as a source for the R-R Kalb-Ramond field $H_2$.
This dual string is
a natural candidate for a fundamental string of a strongly coupled
phase of the IIB theory.
Upon toroidal reduction to six dimensions, $SL(2,Z)$ combines with the
$T$-duality group to give an $SL(5,Z)$ symmetry
of the equations of motion. This leads to a large
multiplet of string solutions, which includes both the
fundamental and solitonic type solutions.

\newsec{Discussion and Conclusions}

In order to put our results in proper perspective, it is useful to
draw an analogy with $d=4$ super Yang-Mills theory. For the case of
$N=4$ supersymmetries, the low-energy effective field theory is (at generic
points in
the moduli space)  an
abelian theory, and has an $S$-duality symmetry
which exchanges weak and strong coupling. It has been conjectured \mol\
that this
extends to an {\it exact} symmetry of the theory. If true, the
conjecture implies that there are two microscopic definitions
(which involve different coupling constants) of
every physical theory. The
symmetry would then constrain not just the leading term in the
low-energy effective action, but every term in a derivative expansion of the
exact theory. A key observation \refs{ \osb, \wol, \sentform}
which makes this conjecture plausible is that
the symmetry does extend beyond the low-energy effective action
at least to the
spectrum of stable BPS saturated states.

It is also true, as has recently understood in detail \swit, that for pure
$N=2$ Yang-Mills theory the low-energy effective field theory is (at generic
points in
the moduli space)  an
abelian theory, and has an $S$-duality symmetry
which exchanges weak and strong coupling. However it is not plausible,
and has not been proposed, that this extends to {\it exact} symmetry of the
theory
(although the effective symmetry is nevertheless extremely useful).
Indeed the symmetry can not even be extended to the BPS states because there
are
no spin one monopoles. Thus the $N=2$ theory has an {\it effective}
$S$-duality, in contrast to the $N=4$ theory which may have an {\it exact}
$S$-duality.

In many regards, discussions of string-string duality to date are somewhat like
discussing $S$-duality in $d=4$ Yang-Mills theory without knowing
the spectrum of BPS monopoles. Most of the discussions involve only the
low-energy
effective field theory and do not really distinguish between exact and
effective
duality\foot{An exception to this, emphasized in \wit\ and
implicit in \town, is the observation that a consistent implementation of
the duality of dualities requires properties of
massive string states which go beyond low-energy supergravity.}.  The duals of
the
string winding and momentum states on $T^4$, and in particular the exponential
spectrum of such states which are BPS saturated, involve charged RR states in
the IIA
theory and little is known about the properties  of such states.  One can argue
that such states must exist since one can construct black holes carrying RR
charge,
but detailed questions involving counting of states have not been successfully
addressed from
this point of view.  We will argue that there is a much more concrete picture
of these states which follows from our construction, which therefore
provides strong evidence for an exact duality.

The heterotic soliton of the Type IIA theory on
K3 carries currents which couple to the $U(1)^{24}$ low-energy
gauge fields coming from the Ramond-Ramond (RR) sector of the
IIA theory. If we consider instead of an infinite soliton string
a finite closed loop of string, then at first we might expect that
it can shrink to zero size and disappear since it is topologically
trivial \foot{Although this is not always the case. See for
example the discussion in \wz.} .
However this is not the case if it carries non-zero
RR charge. For example, a closed loop of string where
the coordinates $X^I$ come back to  themselves up to a shift by an element
of the lattice defined by \yid\
when going around the string carries such charge. Since there
are no such charged states in the perturbative spectrum and since
RR charge is conserved, such a string configuration
cannot disappear. It presumably shrinks down to a tiny loop
which is a BPS saturated charged RR state.

Furthermore,  when the
IIA theory is strongly coupled the soliton string becomes weakly
coupled and has a string tension which is much smaller than the
fundamental string tension.  It seems likely then that at strong coupling
in the IIA theory the low-lying spectrum can be obtained
by quantizing the soliton string as a weakly coupled fundamental string.
This clearly leads to the detailed spectrum of BPS saturated, charged
RR states which are required by string-string-duality and  also  to enhanced
gauge symmetries at the special points in the $K3$ moduli space which
map to enhanced symmetry points in the Narain moduli space.
One obvious question that arises in this point of view is whether this
will ``double count'' neutral states such as the graviton, once as excitations
of the fundamental string and once as excitations of the soliton string. We
believe the correct point of view is that these are two dual descriptions
of the same state, analogous to the description of the photon in $N=4$
Yang-Mills theory as either an ``electric'' neutral gauge boson at strong
coupling or a ``magnetic'' neutral gauge boson at weak coupling.

In some sense the case for an exact
six-dimensional string-string duality is even more compelling than that for
an exact $S$-duality in $N=4, ~~D=4$ Yang-Mills. In the latter case the
spectrum of BPS states is
largely fixed by supersymmetry, so there was little room for disagreement with
the predictions of $S$-duality. In the
string case supersymmetry allows infinitely
many possibilities for the spectrum of the solitonic string, so the
emergence of the heterotic string is  quite striking.

The construction dual to the one presented here would be the
construction of a IIA string on $K3$ as a soliton in the heterotic string
on $T^4$. A natural candidate exists for such a solution,
it is again the symmetric fivebrane,
now interpreted as a solution with non-zero gauge fields describing a
Yang-Mills
instanton. However at a generic point in the moduli space of heterotic
strings on $T^4$ the gauge symmetry is $U(1)^{24}$ and there are no finite size
instanton solutions, at least in the field theory limit. At points with
enhanced
gauge symmetry there are finite size instantons but there is no obvious sign of
the
$K3$ factor in the instanton moduli space which would be needed to give the
correct IIA soliton string. This is just a higher-dimensional version of the
conundrum which arises in the study of $H$-monopoles and $S$-duality
in four dimensions where again a $K3$ factor in the instanton moduli space
seems to be required  for the correct counting of states \gau.
 This is perhaps not surprising given the
fact that six-dimensional string-string duality can be
related to four-dimensional $S$-duality. A resolution of this puzzle is likely
to play an important role in a deeper understanding of duality in string
theory.

In closing we mention that the six-dimensional theories have additional
$p$-brane solitons
corresponding to
fivebranes which wrap around cycles of $K3$ or $T^4$ with
dimension less than four. It is of interest to understand how these
solitons behave under various duality transformations, and in
particular which of them become light at strong coupling.  Clearly at
present that
duality can be best understood in those contexts that involve just
strings and not other $p$-brane like solutions. On the other hand
the close relation between six-dimensional string solutions and $p$-brane
solutions
in higher dimensions suggests that we have not yet heard the end
of the story as far as $p$-branes are concerned.
 The precise interplay between
the treatment of the theory in ten and six dimensions and the role,
if any, played by $p$-branes other than strings should be fascinating.
\bigskip
\noindent {\it Note Added: }
After completion of this work, an interesting paper appeared \sentwo\
in which
the external field configuration to which charged excitations of the soliton
described herein give rise was constructed by dualizing the results of \senone.

\bigskip
\centerline{\bf Acknowledgements}\nobreak

We thank J. Gauntlett, E. Martinec and G. Moore for discussion.
The results on six-dimensional string-string duality
in toroidal type II compactifications were mostly
obtained several years ago in the context of ongoing collaboration with
C. Callan, and we thank him for collaboration and discussions at that time.
This work was supported in part
by NSF Grant No.~PHY 91-23780 and DOE Grant No. DOE-91ER40618.
\listrefs
\end